\documentstyle[epsf,epsfig]{article}

\newcommand{\be}{\begin{eqnarray}}
\newcommand{\ee}{\end{eqnarray}}

\newcommand{\IA}{$\bar I I \,\,$ } 
\begin{document} 

\begin{center}
{\large\bf Toward the Semiclassical Theory \\
of     High Energy Heavy Ion Collisions
} \\
  E.V.Shuryak\\
{Department of Physics and Astronomy,\\ State University of New York,
Stony Brook NY 11794-3800}
\end{center}

\begin{abstract}
Sudden deposition of energy at
 the early stage
of high energy heavy ion collisions 
 makes
virtual  gluon fields real. 
 The same is true for 
 virtual vacuum fields $under$ the topological barrier, excited to
 real states $at$ or $above$ the barrier,   gluomagnetic clusters
of particular structure related to the $sphalerons$ of the
electroweak theory. Semiclassically, these states play the role of
the {\em ``turning points''}. After being produced
 they  explode into
a spherical shell of  coherent  field which then
turn into several outgoing gluons. Furthermore, this explosions
promptly produce quark pairs, as seen from
explicit solution of the Dirac equation.
 The masses of such clusters depend on their size, and 
are expected to peak at $M\sim 3\, GeV$.
After we briefly review those concepts in a non-technical manner,
 we discuss what observable consequences the production of such clusters 
would make in the context of heavy ion collisions, especially
  at the RHIC energies. We  discuss
entropy and especially quark production,
event-by-event
 fluctuations in collective effects like radial and elliptic flows
and $J/\psi$ suppression. Coherent fields and their
geometry increase the jet quenching, and we also
point out the existence of ``explosive edge'' which jump-start
collective effects and may affect unusual phenomena 
seen at RHIC at large $p_t$.
\end{abstract}

\section{Introduction}
\label{sec_intro}

This work\footnote{Its early version exist as a preprint and a
  talk at Quark Matter 2002 \cite{Shu_how}.} consists of two very
  different parts.  Section 2 is 
brief non-technical review of the semiclassical theory of high energy
collisions, covering recent progress described in details in more
technical works. (We think it is really necessary, and helpful
for the readers, to present some  coherent picture first:
 although most  concepts
 are not new they were mostly developed not in the context of QCD, and
certainly not for heavy ion collisions.) Its only new part is section 2.2,
in which the main idea is demonstrated by a simple quantum-mechanical
 toy model, treated without any approximations.

The main body of the paper includes
 applications of these ideas to heavy ion collisions.
In section 3 we discuss why we think they are relevant for the
initial stages of the collisions,
especially at RHIC energies. We continue with discussion of jet
 quenching
and phenomena at the edge of the system in section 4, ending with
possible manifestations of clusters in event-by-event fluctuations
in section 5.  

\subsection{Tunneling and related excitations in QCD}

The semiclassical theory of
the tunneling phenomena in the QCD ground state, associated with topology
structure of the Yang-Mills fields is based on the so called
instanton solutions \cite{BPST,tHooft}. It is  by now
understood in significant detail and has
strong ties to hadronic phenomenology and  lattice studies. These
phenomena are known
to play an important role in chiral symmetry breaking, hadronic
spectroscopy, form-factors etc,
see e.g.\cite{SS_98} for a review. 

The semiclassical approach to
various high-energy
 reactions is much less developed. It had started
 in
the early 1990 \cite{electroweak,DP}, when {\em baryon-number violating}
tunneling in the electroweak theory 
has been actively discussed. 
It waned several years later 
when it became clear that those fascinating phenomena cannot be
experimentally observed.

In a very crude way, one may say that collision of two (virtual) gauge bosons,
W,Z in electroweak theory or gluons in QCD, can produce two type of
objects
shown in  Fig.\ref{fig_almond_diagrams}(b),   a single on-shell gauge
boson or  a turning state cluster .
The former is the so called Lipatov
 vertex,
the main ingredient of the BFKL ladder 
\cite{BFKL} providing high energy asymptotic behavior in pQCD.
The latter is the process to be discussed in detail below.

The first QCD effects of similar origin discussed were
instanton-induced  multi-jet production: the search in
this direction continues at HERA, 
see recent review in \cite{Ringwald}.
However, due to large scale involved and  small-size
instantons, this phenomenon is also associated with  small 
cross sections, and even if the predicted signal
be found it would  not be easy to prove that it is not
due to some perturbative diagrams. 

The situation is different at the so called semi-hard scale
$Q\sim 1-2 \, GeV$, at which tunneling phenomena are
in some cases so large than they simply dominate
the contribution from the perturbative processes. 
Recently it was  suggested that instanton-induced processes
may explain why  the cross section of
 high energy hadronic scattering grows with energy
 \cite{Shu_01,KKL,NSZ}. The reason is prompt production of
specific hadrons (such as the scalar glueballs) and the topological
clusters
to be discussed in this work.
Specific properties of the so called 
{\em ``soft pomeron''} (such as its intercept and the slope)
have been calculated with rather
reasonable results. As unexpected 
 qualitative insights found was a possible reason why there is  no
$odderon$ in this theory \cite{NSZ}). New experimental test
were suggested such as glueball
production
 rate \cite{Shu_01}.

What made the situation qualitatively different now  is a real breakthrough
in understanding of the nature of semiclassical path
including the structure of the turning states and its real (Minkowskian) evolution. 
We will describe those in the next section.

\subsection{Confronting the RHIC puzzles}
  Let us now switch to brief overview of heavy ion physics.
Relativistic Heavy Ion Collider
(RHIC) project have been completed in 2000 and made its first
full-scale run with AuAu at full energy in 2001. The aim of it was
to detect   a transition of the
QCD vacuum to qualitatively new phase,
a deconfined and chirally symmetric
 {\em Quark-Gluon Plasma} (QGP) \cite{Shu_80}.
Among the predicted signals for it were robust collective phenomena
known as radial and elliptic flows \cite{hydro}.
Already the first data from RHIC have quantitatively confirm these
predictions, with equation of state consistent with that expected from QGP. 
Excellent agreement of all spectra goes till $p_t\sim 2\, GeV$, or for
about 99.5 percent of all particles.  (At higher $p_t$ one observed
strong
deviations from the naive parton model: we will return to this in
section \ref{sec_intro_quench}). Studies of correlations between $\pi
K$ or $KN$ agree with significant time difference $\sim 5-6 fm/c$ 
 between emission
of $K$ and $\pi,N$. Event-by-event 
fluctuations are small and roughly consistent
with thermal ones at freezeout. Composition of all secondaries agrees
well
with thermal production rates.
All of it confirm that strong interaction in the
system starts at time $\sim 1/2 \, fm/c$ and ends at
 $\sim 10-15 \, fm/c$,  which means that the QGP is actually 
created at RHIC.

\begin{figure}
 \begin{minipage}[c]{7.cm}
 \includegraphics[width=5cm]{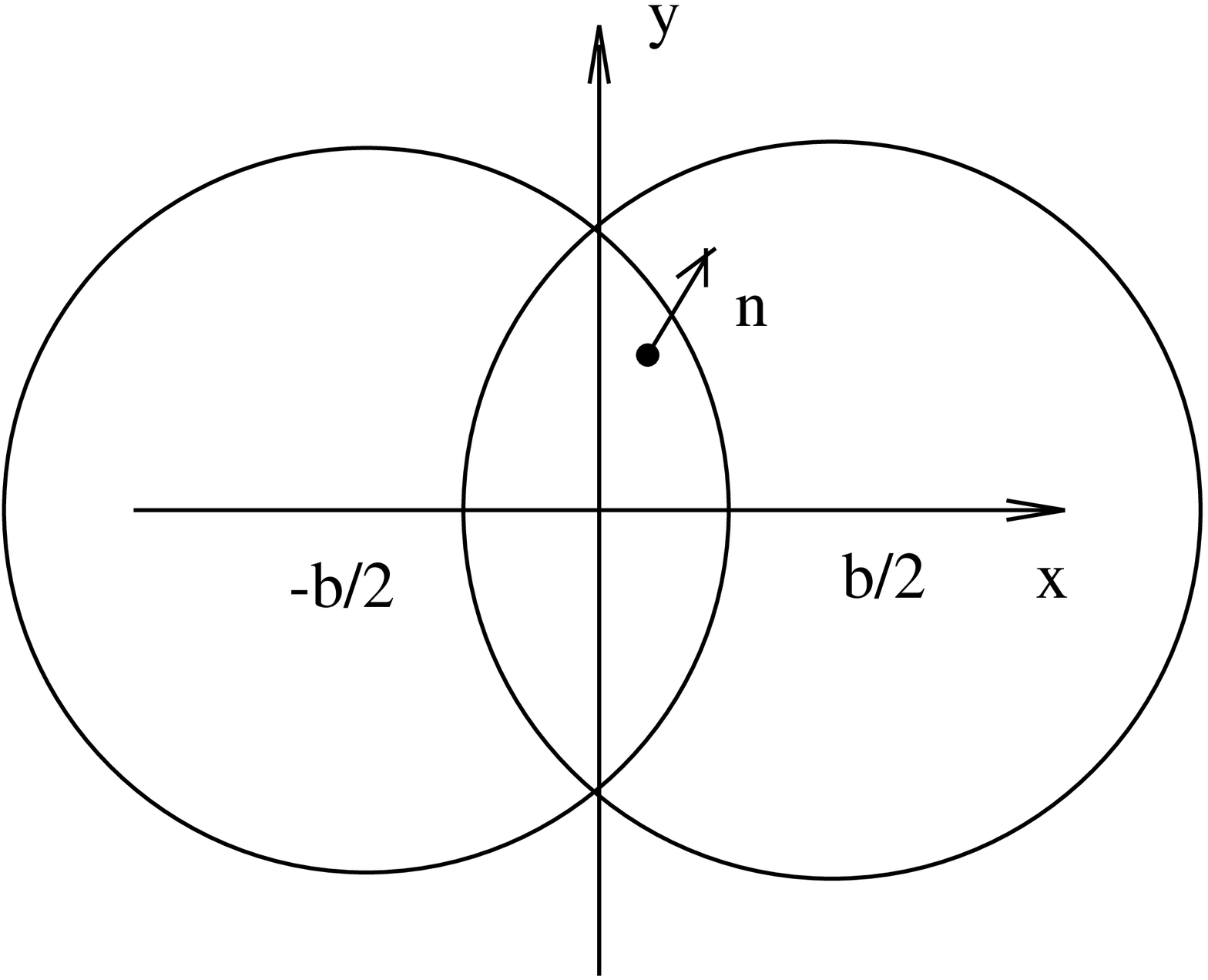} 
 \end{minipage}
    \begin{minipage}[c]{3cm}
    \includegraphics[width=2.6cm]{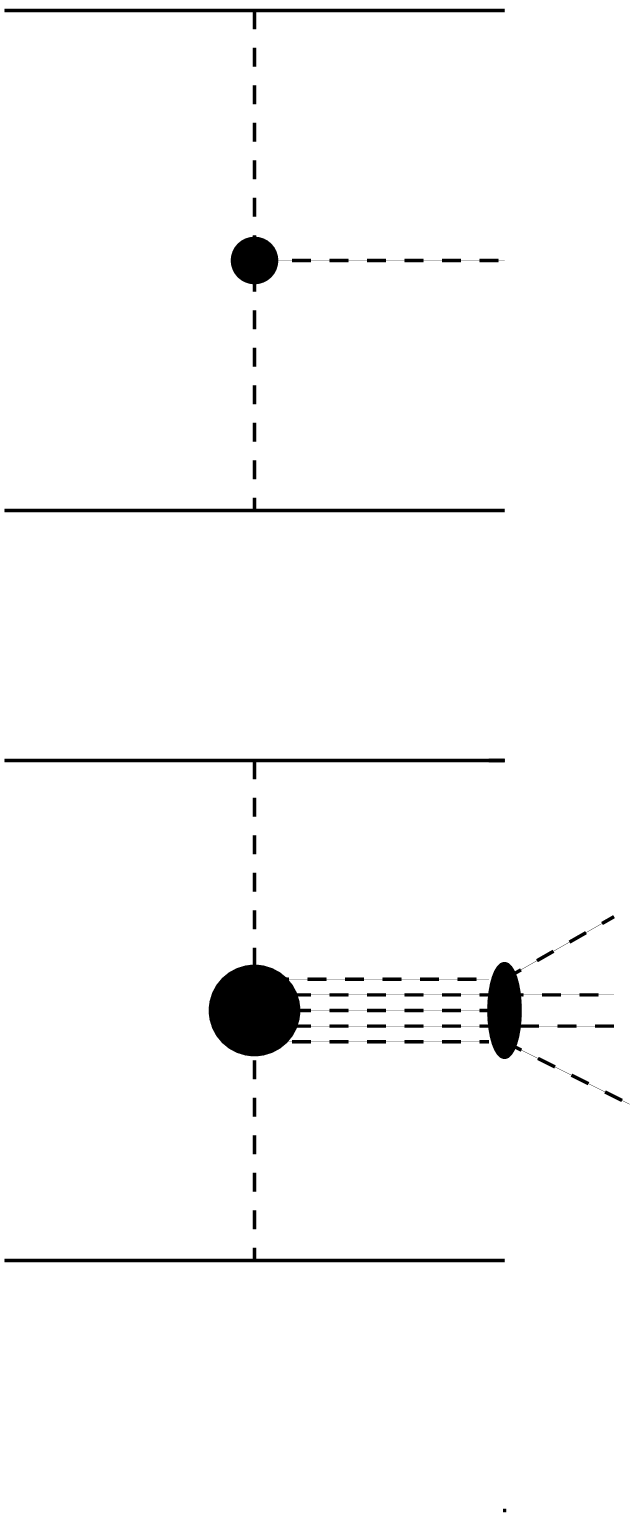}
    \end{minipage}
\caption{\label{fig_almond_diagrams} 
(a) The almond-shape overlap region for the non-central nuclear
  collisions
with impact parameter $b$. (b)    
Two generic vertices producing either a
 physical gluon or  a topological cluster 
 from two colliding virtual gluons in a high energy collision. 
 }
   \end{figure}

Let me explain a bit how the lower limit on this time was obtained.
The elliptic flow\footnote{A special role of elliptic flow
stems from the fact that  it
allows  to separate the initial-state (e.g. parton re-scattering)
 from the final-state interactions, since the direction of event
impact parameter is unknown prior to the collision.} 
is characterized by the parameter $v_2=<cos(2\phi)>$. It is
is present only for non-central collisions, in which the
overlap region of two nuclei have an almond-type shape
shown in Fig.\ref{fig_almond_diagrams}(a).
As we already mentioned in the introduction, hydrodynamics
describes $v_2(p_t)$ well into the tails of
particle spectra, up to  $p_t \sim 2 \, GeV$.
 At the same time  other
approaches -- such as  string-based or mini-jet-based ones --
 have problems reproducing growing  $v_2(p_t)$.

  The crucial point is that
short
initiation time 
 $t_i\sim 0.5-1 \, fm/c$ is absolutely necessary
in order to account for
large observed $v_2$. If  a period
of ``free streaming'' of partons  would exist for longer time, 
 it is impossible to recover ellipticity of the  shape.
 Moreover,  following the observed impact parameter
dependence of $v_2$ one finds that it peaks at large impact parameter
 $b\sim 8-10\, fm$, for which the width of the almond in   
 Fig.\ref{fig_almond_diagrams}(a) is very small.
And still all the hydro predictions apparently work, forcing us
 to
conclude that the mean free path and equilibration time are much
 shorter than this width.

The main issue now is to understand $how$ it happens, what is the
dynamics  responsible for QGP formation in a very such short time.
Let me put this question as the first in 
 the list of the {\em  RHIC puzzles} which remains basically unexplained
\begin{itemize}
\item  Early formation of significant pressure leading to 
explosive collective behavior
\item Large deficit of large-$p_t$ hadron yields, or jet quenching
\item Large (and approximately $p_t$ independent) azimuthal asymmetry at
 large-$p_t$
\item Large  baryon/meson ratio  at  large-$p_t$,  much larger than
in the usual 
jet fragmentation
\end{itemize}
Those are only questions directly related to data, but there are many
more issues of crucial importance which we have to answer, such as:
at what time quarks appear, is excited matter produced gluonic
or closer to equilibrium QGP?

\subsection{Theory overview}

 To put these ideas in proper context, recall that  initial stages of high
energy
heavy ion collisions have been for a long time associated with
 re-scattering of partons with momenta about 1-2
GeV (later called mini-jets)  considered already in \cite{Shu_80}.
 The  
parton cascade model by Geiger and Muller \cite{GeigerMuller} have added
``branching''
of virtual partons, or bremsstrahlung. Important role of  pQCD
processes of
2-to-n type has been discussed
in \cite{SX}.
Recent development along the pQCD line \cite{BMSS}
included account for Landau-Pomeranchuck-Migdal effect and other
improvements. Although perturbative equilibration
was found possible, its rate  is very slow, and basically no
collectivity
was predicted by the mini-jet models like HIJING. Furthermore, 
 it was nicely
quantified in \cite{GM}
by how much pQCD-based scenario with 1-2 \, GeV cutoff misses
what is needed: in order to
reproduce the elliptic flow
by a parton cascade   the product
of the {\em gluon density\footnote{Note however, that the multiplicity of partons
cannot exceed that of final hadrons, because of quite
fundamental limitation: the {\em entropy can never
decrease}.} times the cross section} should be
 increased by a factor of about 80. 
So, naive
extrapolation of pQCD to momenta $Q\sim 1 \, \, GeV$  have failed
to reproduce RHIC data. This has created a serious challenge to
the theory.

  Significant attention has been attracted by
 the idea stemming from a possible {\em saturation} \cite{saturation}
phenomenon at small x, which get some support by HERA
data on deep inelastic scattering.
It was further argued that when, at sufficiently
 small x, the gluon occupation numbers reach the magnitude
$\sim O(1/\alpha_s)>>1$ the $classical$ approach 
to YM field becomes possible
 \cite{McLV}. Such matter was called
the Color Glass Condensate (CGC), and
 its spontaneous ``materialization'' may
  significantly contribute to the QGP formation, as recent numerical
studies using classical Yang-Mills equation  have shown \cite{KV}.

 We will argue in this work that  the CGC is not just transverse
classical field, but a part of it possesses topological properties.
In terms of the filed strength it means that the electric and magnetic
field strength are not orthogonal, $\vec E\vec B$ is substantial,
which leads to quark/chirality production due to chiral anomaly.
We will show that the  energy range of RHIC
 provides an especially good
 window of opportunity to see fascinating
phenomena related to topological tunneling, producing
 a 
 spectacular ``firework'' of exploding
 topological clusters.

Recently  first steps were made toward the understanding
of  the quite different limit,
the  {\em strong coupling limit} of the QGP phase.
Specifically  the so called 't Hooft coupling is
assumed to be large $g^2 N_c>>1$, and indeed in QGP produced at RHIC, with
$T=(1-3)\, T_c$ this seems  to be the case. 
The calculations are done via the famous AdS/CFT
correspondence for $\cal{N}$=4 supersymmetric Yang-Mills theory.
In it the coupling does not run and one can study
 the theory at any value. Let me mention
only one paper of the kind \cite{PSS} focused on shear  viscosity at
high T. Their main result is the following shear viscosity coefficient 
\be \eta={\pi \over 8} N_c^2 T^3\ee
which is significantly smaller than the weak coupling results
$\eta \sim  N_c^2 T^3/\alpha_s^2log(\alpha_s)$,
\cite{AMY}. 
If confirmed that these results can be
used for QCD, it would mean that the mean free path is
so short that 
one can use ideal hydrodynamics, but probably not a parton cascade.

\section{ Quantum Mechanics of the Yang-Mills fields   } 
\label{sec_QM}

\subsection{Topological coordinates and the tunneling paths}
In this section we make  brief review of the general setting,
aimed at non-specialists and
 explaining the main ideas and vocabulary used.

 Quantization of YM fields is simpler in the $A_0=0$ gauge,
in which the canonical momentum  is electric field 
$``\vec P''={d\vec A \over dt}=\vec E$ and the electric part
of the energy $\vec E^2$ is identified as the kinetic term. The nonlinear
magnetic term is identified as a
 potential energy: so schematically YM  can be viewed
as
many coupled non-linear oscillators with a potential of the type
$\vec B^2 \sim (A^2+A^3+A^4)$.

\begin{figure}[ht!]
\begin{center}
      \includegraphics[width=2.6in]{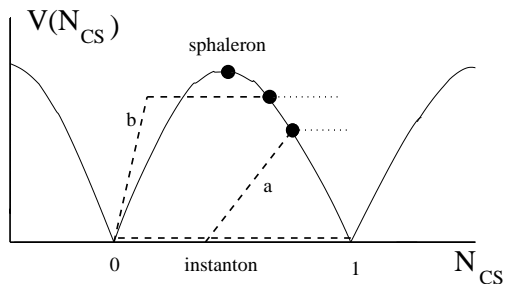}
\end{center}
  \caption{
   \label{fig_potential}
Schematic plot of the energy of Yang-Mills field versus
the Chern-Simons number $N_{cs}$. It is a periodic
function, with zeros at integer points. The $instanton$ (shown by the
lowest horizontal
dashed line) is a transition between such points. However if some
nonzero
 energy
is deposited into the process during transition, the virtual paths
 (the dashed lines) emerges from the barrier,
via the  {\em turning points} (black circles). The later
real time 
motion outside the barrier (shown by horizontal dotted lines)
conserves energy, as the driving force is switched off.
The maximal cross section corresponds to the transition around
the top of the barrier,  the
$sphaleron$.
 }
\end{figure}

A specific combination of $A_\mu$, called  the
{\em Chern-Simons number}
\be \label{eqn_CS}
N_{CS}=\int d^3x K_0\ee
is a special coordinate
related to the so called topological current $K_\mu$. 
\be K_\mu=-{1 \over 32 \pi^2} \epsilon^{\mu\nu\rho\sigma}
({\cal G}^a_{\nu\rho} {\cal A}^a_\sigma -
{g \over 3}\epsilon^{abc} {\cal A}^a_\nu  {\cal A}^b_\rho  {\cal A}^c_\sigma)
\ee 
The {\em potential energy} of the Yang-Mills field versus this coordinate
is  schematically shown
in  Fig.\ref{fig_potential}:  
 it is the periodic
function, with zeros at all {\em integer} points.
Those are called ``classical vacua'', they have zero fields but
non-zero and topologically distinct $A^a_m$. The QCD vacuum is a
quantum superposition
of ground states near all of those classical vacua with the same
amplitude\footnote{Other states exist also, but in those strong interaction
would have a non-zero CP violation excluded experimentally.}.

  The natural
 language to describe
 {\em quantum mechanical tunneling} is Feynman path
integral formulation. The tunneling can be described by specific
 paths,  which start at one minimum of the potential
 and end up
in another.
The path with the minimal action is called
 the $instanton$, it satisfies $classical$ equations
of motion in Euclidean time and dominates the Feynman integral.
The line on the bottom of that figure
is such an  {\em instanton} (shown by the
lowest
dashed line):
it corresponds to the {\em zero} energy solution.
 The corresponding (Minkowskian) action $S_{inst}$
is imaginary and thus the tunneling probability is $\sim exp(-|S_{inst}|)$.
All this has been known since mid-70's.

 In this paper we focus on other paths shown in this Figure.
At the moment of high energy
collisions 
{\em a sudden localization} of all quantum coordinates including
the topological one
takes place. Although their values remain about the same
as they were prior to the collision
moment, the system 
 suddenly get
 placed {\em at or above the barrier} (
this case is shown by the dashed line (a) in Fig.\ref{fig_potential}).
Similar phenomenon is well known perturbatively: the  partons
(virtual field harmonics of the target or projectile) after collisions
 becomes real outgoing radiation
.
The same phenomenon  happens for non-perturbative virtual fields
  as
well: therefore some of the turning states we will discussed below are 
 {\em remnants of the
interrupted instantons}, already present in the vacuum.

Another possibility  (
 shown by the dashed line (b) in Fig.\ref{fig_potential})
is that a system at the collision moment is $not$
under barrier, but becomes able to tunnel through it because
it gets excited enough. The corresponding amplitude would include 
 exponential factors (as all tunneling amplitudes), but not some of the
pre-exponential ones (such as quark condensates) which are included
in the
``instanton density'' of the QCD vacuum.
Whatever way the system is driven, it emerges from under the barrier
 via what we will call  {\em ``a turning state''}, 
familiar from WKB semiclassical method in quantum mechanics.

  The gluonic clusters to be discussed in this work are such turning
states of the QCD paths. Those are relatives of the so called
$sphaleron$\footnote{``Ready to fall'' in Greek, according to
Klinkhammer and Manton \cite{Manton}.} solution of electroweak theory.
This is a point where 
the path crosses the barrier and total energy is equal to only
potential one.
 From there starts the
 real time 
motion outside the barrier. 
Here the action is real and $|e^{iS}|=1$. That means  that
whatever happens  at this Minkowski stage has the probability 1 and cannot
 affect the total cross section of the process:
this part of the path is only needed for understanding of the
properties of the
final state.


\subsection{Exciting a  quantum system from under the barrier }
\label{sec_doublewell_sph}

\begin{figure}[t]
\centering
\includegraphics[width=4cm]{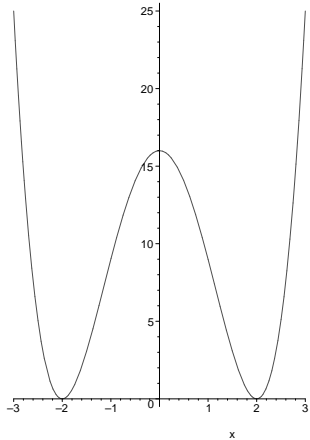} \hskip 1cm
\caption{\label{fig_doublewell} 
 The double-well potential used.   
}
\end{figure}

  Let me supplement the theoretical ideas in the field theory context
  by a simple quantum-mechanical example of related phenomenon, 
which anybody can test
without much difficulty. 
The setting can be any problem with a barrier and tunneling, for
  example the very
often used {\em double well potential}  
\be V = \lambda \,(x^{2} - \mathit{f
}^{2})^{2} \ee
in which a particle of m=1 is placed.
For  the parameters $\lambda=1,f=2$
it is plotted in Fig.\ref{fig_doublewell}. The value of
$f$ is selected since the maximum of the potential $V(0)=16$ 
(the ``sphaleron mass'' of this problem) makes about the same number
of oscillator quanta in a well as in QCD applications.

  The ground state wave function is well familiar to everybody,
it has two maxima corresponding to both wells, at $x\approx\pm f$,
with relatively small probability below the barrier, at $x\sim 0$.

The question we would like to ask is what happens if one rapidly
localizes
the quantum particle under the barrier. 
One may view it pictorially as a narrow beam of particles
localized near x=0 and able to excite the system.
More specifically, we are
interested in the final states arising from such an experiment.

To answer those questions, let us introduce an external periodic perturbation
acting on the system
\be \delta V(x,t)=f(x)exp[-it\omega]\ee
with $f(x)$ well localized under the barrier, 
at $x\sim 0$. The specific shape of $f(x)$ does not matter as soon as
it does not extends to the wells, where the ordinary oscillation
quanta
(analogs of gluons) can be excited. I have
used several of them and will show results for $f=exp(-4 x^2)$ and 
$f=1,|x|<.2;f=0,|x|>.2 $.

The time dependence can be tuned to excite the n-th level $\omega=(E_n-E_0)$,
and then the excitation probability
\be P_n\sim |<0|f(x)|n>|^2\ee
be calculated directly from numerically calculated wave functions and
energies. For even excitation functions $f(x)$ used,
 only even levels  n=2,4 etc can be excited.

\begin{figure}[h]
\centering
\includegraphics[width=4.9cm]{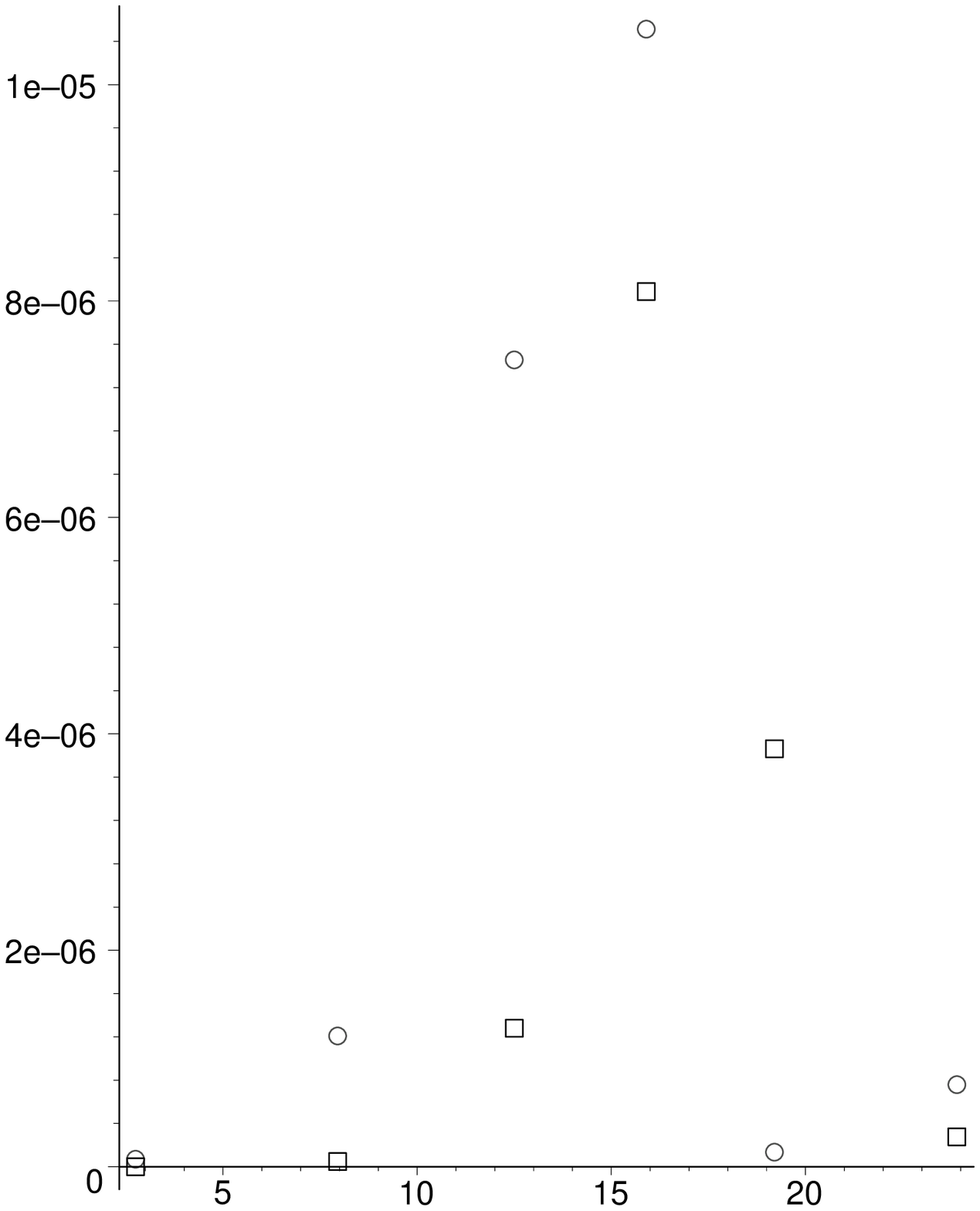} \hskip .2cm 
\includegraphics[width=4.9cm]{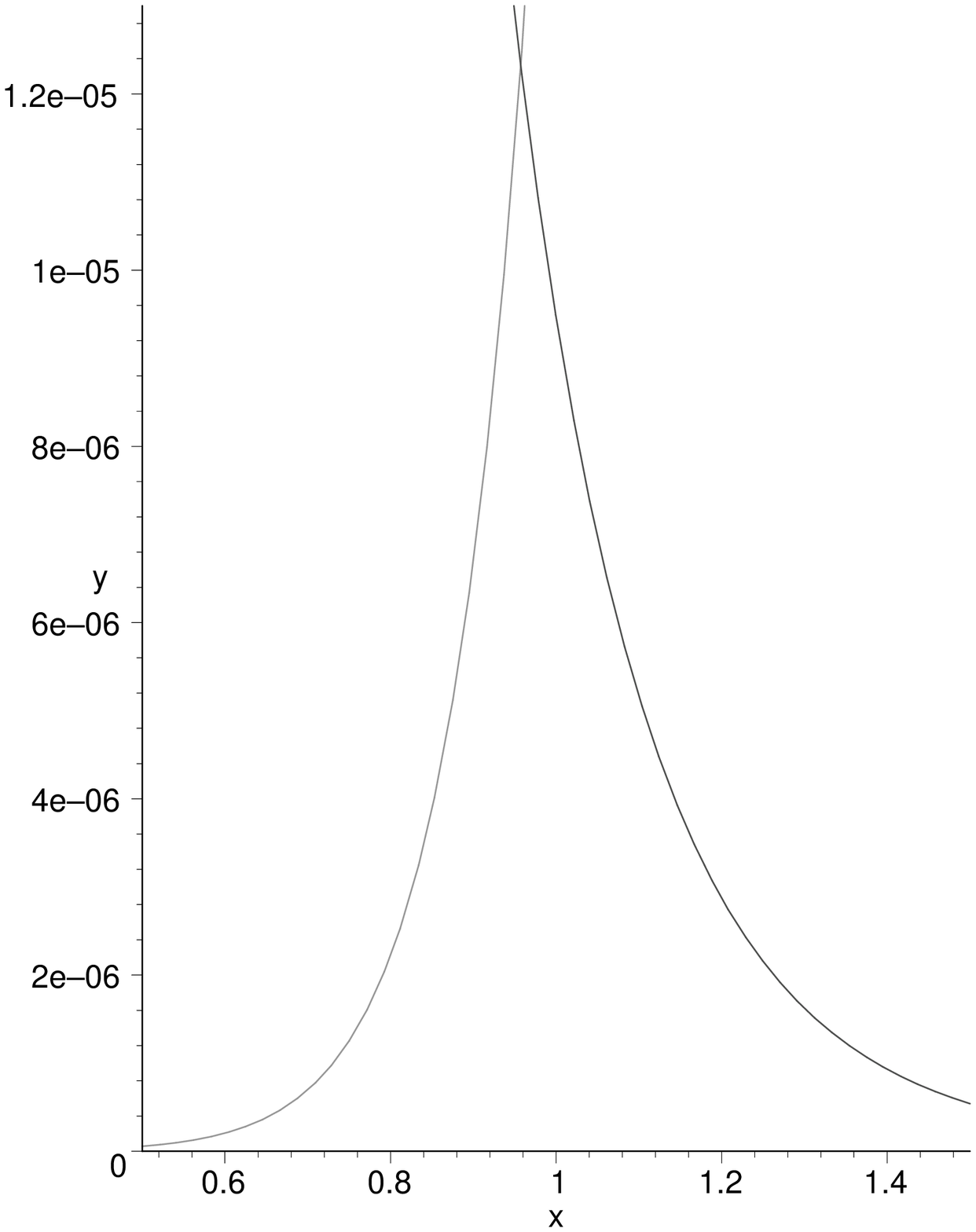}
\caption{\label{fig_crosssection}
(a) The excitation probability $P_n$ of the double-well system
versus the excitation energy. Two sets of points are for two
excitation functions $f(x)$ mentioned in the text.
Note that the peak  excitation energy
corresponds to the maximum of the potential, $V\approx 16$.
(b) The inclusive gluon cross section of the process $gg\rightarrow
sphaleron \rightarrow g+...$
 versus the
energy, in units of the sphaleron mass $x=Q/M_s$, from \cite{JSZ}.}
\end{figure}

The result of the calculation is shown in the Fig.\ref{fig_crosssection}(a).
Note the strong peaking at the excitation equal to the ``sphaleron
mass'' $\omega\approx 16$, indicating that the main 
 final state is sitting $at$ the barrier top. 
 The reasons for the peak are as follows. For energies much less than
 $V(0)$
both the ground and the final wave functions are small under the barrier. For
energies
well above it $\psi_n(x)$ 
 is not small but rapidly oscillating, so that its overlap
with $f(x)\psi_0(x)$ is small. 

Can one calculate the cross section
for sphaleron production semiclassically in the
field theory context?

  We have already mentioned in the Introduction that 
the calculation of the cross section from first principles
is not yet available. The
original semiclassical approach
with vacuum (undeformed) instantons
was pioneered by  Ringwald and Espinosa \cite{electroweak},
who noticed that multi-gluon production is more (not less as in pQCD)
probable than few-gluon one. Unfortunately it 
 is good only at low
energies
of partonic sub-collision, much below the sphaleron mass
(10 TeV in electroweak theory and
 3 GeV in QCD). The problem to find a general semi-classical
 answer was
known as
 the so called {\em holy grail } problem. 
Three methods toward its solution have been proposed

(i){\em Unitarization} of the multi-gluon amplitude when it
becomes strong was first suggested by
Zakharov and worked out by Shifman and Maggiore
\cite{electroweak}. Basically
one can
treat a sphaleron as a resonance, and even the resulting 
expression for the cross sections in \cite{NSZ}
looks similar to Breit-Wigner formula. This is the most worked
out approach, but it still cannot guarantee the parametrically
accurate
 numerical values of the
cross sections.

(ii){\em Landau method with singular instantons} was applied by
Diakonov and Petrov \cite{DP} (following some earlier works
which are cited there)
who were able to find the opposite limit of high energies. It follows
from the comparison of the two limits, that the peak is 
indeed very close to the sphaleron mass, and the cross section is
very close to be {\em first order} in instanton diluteness, not the
second order as the initial probability. Unfortunately they were not
able to find the solution at intermediate times which would provide 
the turning points of this approach.

(iii){\em Classical solution on the complex time plane} \cite{Rubakovetal}
is another possible direction, in which a zig-zag shaped path in 
complex time includes classical evolution and tunneling in one common
solution.
Unfortunately, this interesting idea also has not been fully
implemented,
 even for toy
models
with only scalar fields considered in this paper. 

It would not be possible to describe here those approaches in any more
detail.
Let me just show the figure Fig.\ref{fig_crosssection} 
from the paper \cite{JSZ}: it shows the low energy and high-energy
approximations for gluon production, calculated following ``Landau
method''
mentioned above. Note strong resemblance to the quantum-mechanical
excitation curves shown in  the Fig.\ref{fig_crosssection}(a). 

\subsection{The turning states}
\label{sec_turning_states}
  Static gluonic turning states has been studied in  
  a recent work \cite{OCS} using two different approaches, and 
their structure has been determined. Without technical details, we
give here some of their properties.

  The first part of that work  used
 the   \IA   configurations. 
 Those can be seen in two different ways. The traditional one
is that such paths 
describe  virtual processes in which the path goes under the
 barrier but eventually ends up in the original minimum, {\em without}
tunneling. 
 Another view ( adopted in \cite{OCS}) is that
such paths would rather be a time history repeated twice,
with positive and negative times being mirror images of each other.
If so, it can be seen as 
 the { $probability$} (rather than the amplitude) 
 of the vacuum excitation by some external current
\be P\sim |<0|j_{ext}|turning \, state>|^2 \ee
The amplitude and its conjugate meet in the middle,
which we will describe by the t=0 plane, also known as
 the ``unitarity cut''.
This is the time moment when the {\em turning states} are born
and released into real (Minkowski) space-time.
 
 We will proceed directly to the second method, constrained minimization.
Classical Yang-Mills
is {\em scale invariant}, its energy can always be changed
by re-scaling of the coordinates. However,
 the  {\em energy  times the r.m.s. radius} $E*R$, in which R can be defined as
\be \label{eqn_radius}
  R^2= {\int d^3r r^2 B^2 \over \int d^3r B^2}
\ee
is invariant under scale transformation.

The corresponding turning states can thus be  obtained \cite{OCS}
from the {\em minimization of the
potential energy of a static Yang-Mills fields}, consistent with two
appropriate constraints:\\
{\em 
 (i) fixed value of (corrected) Chern-Simons number $N_{CS}$ (\ref{eqn_CS}).\\
(ii) fixed value of the r.m.s. size R (\ref{eqn_radius}).
}\\ To find those one should 
 search for the minimum
of the following functional
\be 
E_{eff}={1\over 2}\int B^2_md^3x+ {1\over\rho^2}R^2(A_\mu)+\kappa
 N_{CS}(A_\mu)\ee
 where
$1/\rho^2,\kappa$ are two Lagrange multipliers. 
Although these two terms append YM equations and make it more
complicated,
an {\em analytical} solution is found. Skipping  the details, let
me only say that it is a well-localized magnetic ball, with $\vec B^2$
depending on the radial coordinate only. It is a SU(2) object, which
means that it has 3 (out of 8) gluonic magnetic fields, with their
field lines being closed in circles and rotating around axes x,y,z
respectively.

The total  energy of it times the size depend on $kappa$ as follows
\be E\rho=3\pi^2(1-\kappa^2)^2/g^2\ee
 while the Chern-Simons number is
\be\tilde{N}_{CS}={\rm sign}(\kappa)(1-|\kappa|)^2(2+|\kappa|)/4\ee
Eliminating $\kappa$, one finds the
profile of the topological potential barrier shown in Fig.\ref{fig_potential}.
Its maximum (to be called the YM sphaleron) 
corresponds to $\kappa=0$,  its energy
is $3\pi^2/(g^2\rho)$: the true potential however is exactly symmetric
around $N_{cs}=1/2$.


\subsection{Explosion of the turning states }
\label{sec_explosion}
After one knows the turning states, we can study their fate
by solving the YM equation \cite{OCS} starting from them.
Of course, when field become weak enough those equations are no longer
valid. We return to the question where it happens in section 
\ref{sec_scaleQGP}.

The first study of the kind
 has been made a decade ago in electroweak theory \cite{sphaleron_decay}
for the sphaleron, where it has been found that it decays in about 51
W,Z,H quanta. 
The difference  in QCD is there are no Higgs scalar and its non-zero VEVs, so
gluons
are  massless. This makes the process much more {\em explosive}
because
all harmonics with different momenta move together, with the speed of light.
As described in \cite{OCS}, we had
solved it both {\em numerically}  and {\em analytically}
(based on work by Luescher and Schekhter \cite{Lusch_etc}). If one uses notations
introduced by Witten (accidentally in the same year, 1977)  for spherical YM, it makes it most
 elegant. 4 components of the potentials can be written as
\be{\cal A}^a_j=A(r,t)\Theta^a_j + B(r,t) \Pi^a_j + C(r,t)\Sigma^a_j\\
{\cal A}^a_0= D(r,t) \frac{x^a}{r}
\ee
with
\be
\Theta^a_j=\frac{\epsilon_{jam}x^m}{r}, \qquad \Pi^a_j =
\delta_{aj}-\frac{x_ax_j}{r^2}, \qquad \Sigma^a_j = \frac{x_ax_j}{r^2}
\ee
and functions $A, B, C, \mbox{and} D$ can be thought of as
($r, t$ dependent)
Abelian gauge ($A_{\mu=0,1}$) and Higgs ($\phi, \alpha$)
field on hyperboloid 
\be
A=\frac{1+\phi\sin\alpha}{r}, \quad B=\frac{\phi\cos\alpha}{r},\quad
C=A_1,\quad D=A_0.
\ee
with the action
\be
S=\frac{1}{4g^2}\int d^3x dt[({\cal B}^a_j)^2 - ({\cal E}^a_j)^2]=
\nonumber \\ 
4\pi\int dr dt
((\partial_\mu\phi)^2+\phi^2(\partial_\mu-a_\mu)^2
+ \frac{(1-\phi^2)^2}{2r^2}  \\ 
-\frac{r^2}{2}(\partial_0A_1-\partial_1A_0)^2) ) \nonumber
\ee

Omitting all details, let us show directly 
the promised spherical shell at large times.
Those have the following energy density
\be
4\pi r^2 e(r,t) = \frac{8\pi}{g^2\rho^2}(1-\kappa^2)^2
\left(\frac{\rho^2}{\rho^2+(r-t)^2}\right)^3
\ee

Of course, at large times the field becomes weak field which can be
decomposed into gluons:
the Fourier transform of the fields
provides the energy distribution of the resulting gluons.
Numerical studies of the problem has been reported in 
\cite{OCS} as well. Those are naturally more flexible than analytic
 and allows for
more realistic initial shape of instantons and sphalerons, with
exponential (rather than power-like) tails of the fields at
large distances\footnote{The phenomenological reasons for
exponential rather than power instanton tail
are discussed e.g. in \cite{COS}.
}. 

%

Alternative derivation of the same explosive solution, using specific
conformal mapping from the d=4 spherically symmetric Euclidean
solution
has been found in \cite{JSZ}.
  
  Now, what about quark pair production? Again, 
  this problem much debated in literature on electroweak
theory in 1980's but  not yet completely solved.
In its general form, the issue is to derive an analog of the index
theorem for the case when there are outgoing fields: it suppose to
 tell us how many fermionic levels have crossed zero (and are 
produced) based on some topological properties of the gauge field
alone. The usual form of it, involving a change in the Chern-Simons
number, seem to be
 obviously incorrect, because its variation is in general (and
specifically for the time-dependent solutions we speak about) not an integer.

 Further progress  in this direction was achieved by
derivation of the explicit solution to the Dirac eqn in the background of
exploding sphalerons  \cite{SZ_quarks}. This solution 
starts with a static zero energy solution for the YM sphalerons,
and then 
shows how the quarks 
 get accelerated by the electric field
into the finite-energy spectrum of emitted quarks. The energy spectrum
of the outgoing quarks have been found to be rather simple
\be
{\bf n}_R (k) = \frac {4\pi\,k^2}{2k}
 \,|{\bf q}_R^\dagger(\vec k )|^2 = \rho\,(2k\rho)^2\,e^{-2k\,\rho}\,\,.
\label{Z26}
\ee
The distribution integrates to exactly one produced quark.
It is close to  Plank spectrum with the effective
temperature
$T=1/(2\rho)$, which is about 300 MeV for a standard $\rho=1/3 \, fm$.
Accidentally, this is close to the initial temperature of quark-gluon
plasma in the RHIC energy domain.

 This phenomenon leads to production of quarks from 0  up to  the whole
  set of light quark pairs,  $\bar u u  \bar d d \bar s s$. 
 So
 tentatively we estimate the yield of partons per cluster
to be about 3 gluons and up to 6 quarks+antiquarks.

\section{Initial stage of the High Energy Heavy Ion Collisions}
 \subsection{Excitation of the QCD vacuum into the QGP }
\label{sec_inst}

  We have briefly reviewed recent theoretical literature: now
we discuss more phenomenological issues and the numbers involved.
Let us start with the question:
 what is the size $\rho$ and consequently the mass
of the QCD topological clusters excited?

The idea of sudden excitation implies that all coordinates
of the system are not changed. Size is one of the coordinates, and so
it is natural to assume that the cluster size is approximately the
same as the mean radii of the instantons
populating the QCD vacuum. Using a value  $\rho=1/3$ fm \cite{Shu_82},
which has passed multiple lattice and phenomenological tests, one
obtains the cluster mass of about 3 GeV. (For comparison, the mass
of electroweak sphalerons is about 10-15  TeV.)

Let me remind some basic facts about the QCD vacuum and instantons.
Any  quantum-mechanical
tunneling lowers the ground state energy, and in QCD it is also the
case\footnote{In the presence of fermions this statement is not   
generally true. For examples, super-symmetric extensions of both
quantum mechanics and QCD have {\em positive} shift of the ground
state energy, and for QCD with large number of flavors $N_f$ it was
suggested
that the result even oscillates with $N_f$ \cite{SV_largeNf}.}.
The
 instanton contribution
to the {\em ground state energy} $\epsilon_{vac}$ 
  is calculated in the models and
on the lattice: see \cite{SS_98} for details.
A general expression for the non-perturbative
 shift of the {\it  QCD vacuum energy density} 
is known as the scale anomaly relation\footnote{Although this is often referred to as a ``bag term'',
it is actually 10-20 times the value of the bag constant from the fit
of MIT bag model \cite{Shu_78_conf}.} 
\be \epsilon_{vac} =-{b \over 128 \pi^2}<0|(gG^a_{\mu\nu})^2|0> \ee
where $b=11 N_c/3-2N_f/3$ is the usual coefficient of the beta function
of QCD, and the matrix element here is known as the gluon condensate.
Each instanton contributes $32\pi^2/g^2$ to it, so\footnote{We have ignored
corrections
to the action due to instanton interactions and other effects here.}
the instanton  contribution 
\be \label{vac_energy} \epsilon_{vac}^{instantons}=-{b \over
4} n_{inst} 
\approx -(0.5) \, GeV/fm^3 \ee
where numerical value corresponds to
 the phenomenological instanton density \cite{Shu_82} $n_{inst}=1 fm^{-4}$.
Note that this energy density \cite{Shu_78_conf} is 3 times higher than the weight of the
 nuclear matter  $m_n n_0\sim .15 \, GeV/fm^3$.

One may explain that as follows:
 we  live in a kind of a 
superconducting phase, and only by colliding
heavy ions we can produce a tiny fireball 
 of the ``normal" phase, QGP, which has its
ground state  higher. In order to produce
it, we should  both (i) create  thermally excited quarks and gluons and
(ii) ``melt" this vacuum energy\footnote{
More accurately, at $T=T_c$ instantons do not go away but restructure
into pairs \cite{SS_98}. They are however get suppressed at T relevant
for RHIC conditions due to
Debye-like screening. }.

The crucial point we now emphasize is
 that  this amount of energy
density
is only the {\em lower bound} on the actual energy one $must$ spend
in order 
to erase the vacuum fields from QGP. This value would correspond
to the {\em adiabatically slow}
excitation process, in which it would be
equal to the work  $W=\int p dV$ against the vacuum pressure 
 $p=-\epsilon_{vac}$. 
In the high energy collisions we are
closer to the opposite limit of {\em instantaneous}
excitation, so the  excitation energy is  higher.
As a simple example, consider a volume V instantaneously ``shocked''
so that all instanton sufficiently closed to that moment are
 ``disrupted'' and the
Yang-Mills field are found with $non-integer$
values  of the topological coordinate $N_{CS}$.  
The lowest
energy states in this case is nothing else
but the topological potential we are going to study: it is rather expensive.
The excitation energy density can be estimated\footnote{All instantons
with centers separated in Euclidean time by less than $\rho$ are excited.}
as that  of a gas of ``turning states'' or clusters 
$ \epsilon \sim M_{clusters} n_{instantons}\rho $
where $M_{clusters}\sim 3 \,GeV$
and instanton size $\rho=1/3 \,fm$ and density $n_{instantons}=1 \,
 fm^{-4}$ \cite{Shu_82}. With these numerical values,
 the excitation energy density
is about 1 $GeV/fm^3$, a factor 2 higher
 than in the adiabatic case (\ref{vac_energy}).

Another
 general point: since the instanton vacuum   
is rather dilute $n\rho^4 \sim 10^{-2}$ \cite{Shu_82}, its
instantaneous
excitation naturally leads to rather dilute system of gluonic clusters
(see below).

One more important feature of the QCD vacuum is {\em chiral symmetry
breaking}, which is also believed to be generated by instantons.
Its magnitude is characterized by the quark condensates, which have
the following magnitude\footnote{For completeness: unlike the vacuum
energy this quantity has non-zero anomalous dimension and thus
depends on the normalization scale $\mu$, which is taken to be 1 GeV.} 
\be 
<\bar u u>=<\bar d d>\approx 1.8 \, fm^{-3}, \qquad <\bar s s>\approx 0.8<\bar u u>
\ee
In total it makes about 5 quark-anti-quark $pairs$ per $fm^{-3}$, 
to be compared to about 0.5 valence quarks per $fm^{-3}$ in nuclear matter.
In sudden collisions all condensates should disappear, thus all these
vacuum quarks should become real as well. 

Let us now have a look at heavy ion collisions. For central AuAu
at RHIC energies (which will be our primary example in this
paper) the volume occupied by QGP at its maximum is
$V_{QGP}\approx 1000 fm^3$. Considering now the total energy needed
to kill instantons (non-adiabatically) is of the order of 1 TeV,
and total number of quark pairs from eliminated quark condensates
is in the thousands. This is comparable to total $transverse$
energy and total hadron multiplicity observed: thus the phenomenon
clearly cannot be neglected.

Concluding this subsection, we worn the
reader again that the examples given in it are just to get
some preliminary orientation with the physics and numbers involved.
All of them mentioned came from
 the instanton liquid model \cite{Shu_82}
which was designed to explain the mechanism of the chiral symmetry
breaking
(such as the values of the condensates) 
and masses of the lowest hadronic states.
In this model the  only  instantons which counts are those sufficiently
separated from each other and
contributing significantly to the near-zero eigenvalues
 the quark Dirac operator.
In the applications we consider now -- high energy collisions --
this condition can be lifted.
If so, one finds that the total
 density of {\em instanton-like} topological fluctuations in vacuum
is actually about {\em one order of magnitude larger}. 

\subsection{Why heavy ion collisions and not pp?}
\label{sec_scaleQGP}
  Although topological cluster production is argued to be important
  for
understanding of the hadronic cross sections, and in particular
their slow growth with energy, the main point of this work is that such
phenomena  have a chance to be even more important in
heavy ion collisions. The reason is related with different cutoff
  scales
in vacuum and QGP.

In the QCD vacuum 
 the  non-perturbative effects
generate a ``semi-hard'' or ``substructure scale'' 
   $Q^2 \sim \, 1-2 \, GeV^2$, which is
both
the {\em lower} boundary of pQCD 
as well as the {\em upper} 
 boundary of 
low energy effective approaches (like Nambu-Jona-Lasinio
or Chiral Lagrangians).

 Quite different   pQCD cutoff
is however expected for
 heavy ion collisions.
As  argued over the years (see e.g. \cite{Shu_80})
 the final state is in a Quark-Gluon Plasma  phase of QCD. It is
 {\em qualitatively different} from the QCD
vacuum: 
there is no confinement or chiral symmetry breaking, and instantons
are suppressed as well.
    Therefore the   cut-off in excited matter   is expected
to be  determined by  a plasma-like 
 screening: its 
description in terms of quark and gluon {\em quasi-particles}
becomes natural.
At equilibrium, gluon
effective mass is \cite{Shu_80}
\be M^2_g={g^2T^2 \over 2} ({N_c\over 3}+{N_f\over 6})\ee
Although the scale in question grows with T, 
in the window  $T=(1-3)T_c$ it is  actually 
{\em smaller} than the pQCD cutoff in vacuum.
Lattice thermodynamics data support it, and
fitted quasi-particle masses
(see e .g. \cite{LH})
are 
\be \label{eqn_thermal}  M_T^g\approx .4 \, GeV,\qquad M_T^q\approx .3 \, GeV \ee
Without such low masses one also cannot get
high pressure,
 which is not only seen on the lattice but also  in explosive behavior
of RHIC
collisions.

\begin{figure}[ht!]
\begin{center}
      \includegraphics[width=2.9in]{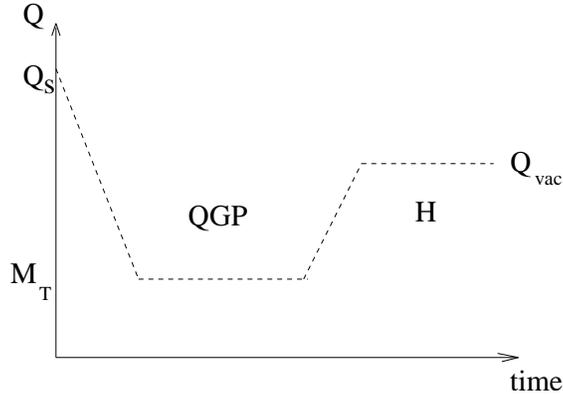}
\end{center}
  \caption[]
  {
   \label{fig_scales}
Schematic plot of the cut-off scales during the evolution
of the system with time. At the collision time=0 the scale
is presumably the saturation scale $Q_s$ in the incoming nuclei,
which grows with the collision energy. Then the cutoff decreases
reaching some nearly constant value in QGP,
the thermal gluon mass $M_T$ (\ref{eqn_thermal})
and stay at this value till it rises again in the mixed phase to its
vacuum value in the hadronic (H) phase $Q_{vac}\sim 1 \, GeV$.
 }
\end{figure}

The schematic picture of scale development with time after collision
is shown in Figure \ref{fig_scales}. 
 A very important consequence of such non-monotonous
behavior of the pQCD cutoff implies then,
 that we can describe gluons (quarks) originating 
from exploding non-perturbative objects  by
classical Yang-Mills (Dirac) eqs with better confidence,
provided those go into QGP in this window of parameters.
This is why heavy ion conditions are so special.

If a topological cluster is produced in hadronic collisions,
its expansion is affected at the vacuum scale $\sim 1 \, GeV$
itself, by confining forces and other instantons. This is above
the mean energy of the spectrum just discussed: it means that
it is impossible to use classical YM description in this case.  

However if it is produced in AA case and gluons/quarks are emitted as
quasi-particles
into deconfined QGP, the matter modification only applies to  momenta
$p< M_g,M_q$ which is till somewhat smaller than
the peak value of  the spectrum resulting from the cluster decay.
So, although such modifications can be substantial, they cannot
destroy most of the classical treatment of the cluster decay.


\subsection{Cluster Production in Heavy Ion Collisions  }
\label{sec_production}

We will argue in this section 
 that at RHIC
cluster production may contribute significantly to
 the total amount of {\em entropy} produced, and, is probably
the leading source of 
promptly produced quark-antiquark
pairs.

As explained in section 2, the semiclassical {\em ab initio} calculation of
the sphaleron excitation is not yet available.
 We will therefore use estimates based on parton-model-style
  phenomenology of all hadronic
collisions, developed  by
G. W. Carter,
D. M. Ostrovsky and myself \cite{COS}. The main idea was to identify
two components of the hh collisions, the color exchanges and
the ``color objects production'', and  deduce the corresponding
cross sections at the partonic level.
 We looked at high energy $NN$, $\pi N$, $\gamma N$, and 
$\gamma\gamma$ cross sections 
 which all $increase$ with energy logarithmically for 
 $\sqrt{s}\sim 100 \, GeV$
\begin{equation}
\sigma_{hh'}(s)= \sigma_{hh'}(s_0) + X_{hh'}\ln(s/s_0)
\end{equation}
We identified the two components mentioned above with these two terms,
respectively, and concentrated on the last (growing) terms.
(In contrast to traditional single-Pomeron fit we do not assume those to
be
proportional to the first terms.) We 
studied whether some universal {\em semi-hard} parton-parton collisions 
can explain all known $X_{hh'}$. Using fitted structure functions of
$N,\pi,\gamma$ and simple scaling -- each gluon can be counted as 2 quarks\footnote{Corresponding to SU(2)
Casimir scaling, appropriate for instanton-induced reactions.} --
we have expressed all of those with only one parameter, the
 value of the {\em qq cross section}. With the fitted
value\footnote{Note that simple parametric estimate for this cross
section,
namely $\pi\rho^2 n_{inst}\rho^4$
 gives the right magnitude.}
\be\sigma_{qq}=1.69\times 10^{-3} fm^2\ee
we got the rising part of cross sections for 4 hadronic reactions
reported in the Table in \cite{COS}. They all agree reasonably well
with the fits for $X_{hh'}$ given by Particle Data Book.
 In \cite{COS} we have also looked at 
the {\em shadowing} corrections, of the second (growing)
component by the first 
 in pp, where the cross section growth 
 at any variable impact parameters is known directly from
 the data. The model qualitatively
describe these data
as well\footnote{As in other works, we found that the average index
of energy growth, 0.08 or so, is actually much reduced by shadowing,
and the true one (seen at large impact parameters) is about twice
this value.}. 

We may now extend our analysis to estimate the number of sphaleron-type
clusters produced from excited instantons in heavy ion collisions. 
For  central $AA$ collisions of two nuclei we use first the
simplest model, one of two spheres with homogeneously distributed partons.
The total parton number is $AN_q$, with  $N_q\approx 12$ being the number of
``effective quarks'' (quarks number plus twice gluons number)
per nucleon\footnote{Of course, the clustering of partons
into ``constituent quarks'' and nucleons increases the number of
collisions, but we will ignore such correlations for now.}.

The total number of $qq$ collisions in this case is easily obtained from 
the following geometric integral:
\begin{eqnarray}
N_{coll} &=& 8\pi\sigma_{qq} n_q^2\int_0^R dr_t r_t \left(R^2-r_t^2\right)
\nonumber\\
&=& 3^{4/3} 2^{-5/3} \pi \sigma_{qq} n_q^2 \left({A N_q \over \pi n_q}
\right)^{4/3},
\end{eqnarray}
where the quark density is determined by the nuclear density to be
$n_q = N_q\times 0.16$ fm$^{-3}$.

With $A=197$ (Au) and the value for the quark-quark cross section
given above we got the
following {\em upper limit} for the
production  of sphaleron-like clusters
at RHIC energies\footnote{We will return to energy dependence
of this production later. This estimate is based on effective number of
partons
integrated from x=0.1 to 1.}
\begin{equation}
N_{objects}  \approx 400 \,,
\end{equation}

(Note that this is the upper limit not because we do not know the
growing part of the hh cross sections but simply because
 the ``objects'' mentioned here can still be either gluons or
colored clusters we discuss, or something else: studies of the cross section alone
cannot tell the difference. For that one should seriously study
the final state corresponding to high-multiplicity hh events:
some ideas what to look at can be found in \cite{Shu_01}.)

This estimate is also naive in the sense
 that nuclear structure functions are
simply assumed to be A times that of the nucleon. One should correct it
for what is known as saturation or  nuclear shadowing.
So far for gluons this effect is unknown, but hopefully will be
 clarified soon. 

\subsection{The Entropy and Quark 
production} 
\label{sec_entropy}
When we mentioned above ``early QGP production'' we meant it in only quite
limited sense, namely that it can drive hydro-like collective effects.
This implies that  that the system is ``optically dense'' $n\sigma
\tau >>1$
so that the  regime is collisional, and also that its EOS is close enough to
$p/\epsilon= 1/3$. 

Many more fundamental questions can be asked such as:
(ii) What is the production history of the {\em total entropy} of the system?\\
(iii) When is {\em the quark part}
 of the QGP  produced?

Certainly AA collisions producing thousands of outgoing mesons
include thousands of quark-anti-quark pairs produced in the process.
However, at AGS/SPS  collision energies those are believed to be generated
 by fragmentation of the QCD strings, or color tubes. This process
takes time at least 1-2 fm/c because pair production contains small
barrier factor.
 
At higher energies of RHIC/LHC, it was long believed
\cite{Shu_80} that the produced highly excited matter remains during
its evolution of several fm/c mostly a ``hot glue'',
with small quark admixture. Various pQCD calculations all concluded the
same \cite{GeigerMuller,BMSS}, as well as those based on classical CGC
 picture \cite{KV}.

 The basic difference between other kind of glue and topological
clusters is that in this case the electric and magnetic
fields have a significant collinear component, the invariant
$\vec E \vec B$ is comparable to $E^2,B^2$. By the chiral anomaly
relation,
this drives chirality and thus quark production. How it happens in
detail {\em during Minkowskian explosion}
can be seen from the dynamical solution discussed in \cite{SZ_quarks}. 
It tells us that if the initial zero mode state of the sphalerons
is populated, at the end those quarks will get physical (positive)
energies $\sim 1/\rho$.

However, it still remains not quite clear what happens with the
fermions
during the first (Euclidean) part of the excitation, which will tell
us
 how much that zero level is populated. In the vacuum, virtual quarks are not
really bound to a particular instantons but are instead
 constantly on the way between
instantons
and anti-instantons. One may think that they equally belong to both
ends of the path, and at the moment of sudden excitation there is a
stochastic process, and the number of produced quark pairs varies
statistically
from 0 to $N_f=3$.   

  Assuming for cluster production a naive (UN-shadowed) upper limit
 discussed above, one gets for
 central $AuAu$ 
collisions at RHIC  $400*3=1200$ gluons
and in average about the same number of quarks+antiquarks.
Together it is comparable to about a half of 
  {\em  the maximal } number of partons, limited by the total
entropy (observed multiplicity
of hadrons) at the end of the collision.  
The mean energy per parton produced via topological
clusters is  about $3 \, GeV/6\sim 1/2 GeV$, a relatively small number.
If those would equilibrate by themselves, its effective temperature
would be only $T\sim 200 \, MeV$. 

One should  compare these numbers with the predictions of other
 approaches such as CGC model.
Based on a numerical solution of the YM equations on transverse
lattice \cite{KV}, at RHIC energies
the saturation scale $Q_s$ was estimated to be 1.3 GeV.
 The transverse energy per quantum 
was found to be $1.66\, Q_s$, resulting in an  effective temperature 
of released gluons $T_{CGC}\approx 1\, GeV$. Thus the gluons from
the  CGC decay are very 
hot, hotter than from the topological clusters. The total number of partons
or entropy is however roughly
 comparable. 

After the initial formation stage is finished, the CGC and the clusters decayed
and QGP is formed, one suppose to obtain 
the equilibrium initial temperature which is known from
 hydro calculations. It is about $T_i\sim 350\, MeV$, or indeed an intermediate
value between  the two temperatures mentioned.

\subsection{Collision Energy, Centrality and Rapidity
Dependence of Entropy Production}

  In fig,\ref{fig_almond_diagrams}(b) we have shown that both perturbative 
gluon production via Lipatov vertex and  the nonperturbative
mechanisms
we discuss may originate mostly from a collision  of two virtual
gluons.
As a result, there are similarities between the two processes, and 
in this subsection we will discuss whether one can use dependence on
any of the parameters (mentioned in its title) to separate the two
mechanisms. In the first approximation, many
 features (such as nuclear shadowing etc) are obviously simply common
 to both of them. 

{\em Rapidity} distribution of both components is about the same,
since both are roughly given by a convolution of two\footnote{We have
ignored in both cases possible multi-parton collisions.
} distribution function\footnote{The distribution function 
$\phi(x,p_t,Q)$ depends on parton transverse momentum: its integral over
$p_t$ is the better known structure function such as xG(x,Q). Here Q
Nstands for a normalization scale defining which partons we speak
about.}.
 As pointed out by Kharzeev and Levin \cite{KL01},
the interplay of (i) the saturation scale
 behaving at small x as 
\be \label{eq_Qsat} Q_s(x)\sim 1/x^\lambda, \,\,\,\, \lambda=0.25-0.3 \ee
 with  
(ii) the saturation conditions itself, leads to
characteristic triangular shape of
the rapidity distribution\footnote{Naturally
this expression also agrees with the
original parton model of Feynman, which corresponds to
 $\lambda=0$ 
and correspondingly  a ``rapidity-independent plateau''.}
\be {d \over dy}\sim exp(-\lambda|y|) \ee 
which agrees well with the RHIC data.

Also we expect the {\em centrality dependence} of both mechanisms
be very similar. Again, following Kharzeev and Levin \cite{KL01}
one can get 
a very good description of it substituting naive number-of-collisions
scaling to saturation with its very slow logarithmic dependence
$\sim 1/\alpha(Q_s(b))$  of density of common small-x gluons produced by
a row ($\sim A^{1/3}$) of nucleons. It is not surprising, since
it is a feature of the wave function of colliding nuclei itself.

The dependence on the {\em collision energy }  however is expected to be
quite different.
The usual parton model, involving convolution of two parton densities
with the power law behavior of the structure function $xG(x)\sim 1/x^\lambda$
would predict also a power growth of the number of partons produced
\be \label{eq_growth}
N(s)\sim s^\lambda \ee
However, the saturation scenario reduces this power to a {\em twice
smaller} power, $\lambda/2$, see  \cite{KL01}. The reason for that
is the energy dependence of the saturation scale itself, \ref{eq_Qsat}.

 So far we have 
implicitly assumed the cluster production scale being fixed by cluster
mass,
or the instanton size in the QCD vacuum, which is some s-independent
fixed scale. We also implied that heavy ion collisions at RHIC
have $Q_s\approx 1 \, GeV$, which is the same as the semi-hard
instanton scale.
If so, the only difference between a single gluon
Lipatov vertex and the instanton-induced cluster production is just 
a factor in the vertex.

  However if we consider energy dependence and ask what happens
at much higher energies, such as LHC ones and beyond, one should
 have a better look at scales involved. We will only provide an estimate,
assuming the saturation scenario, with $Q_s$ increasing as a small
power of s. Simple one gluon production, described by
the Lipatov vertex 
$V_L\sim \alpha_s(Q_s)$ decreases with energy logarithmically.
If one $naively$ substitutes the saturation scale 
into the instanton-induced
amplitude 
\be V_{inst} \sim exp(-S)S^{2N_c}\, \, S=2\pi/\alpha_s(Q_s) \ee 
one finds a power-like decrease\footnote{The contribution
of ``interrupted vacuum instantons'' always remains, but would be
negligible
at very high energies, as it can only provide constant contribution
to multiplicity. }.
It implies \cite{Shu_01} that at  energies  high enough so that
 the saturations scale would be significantly larger than
the ordinary instanton scale, one should  expect suppression of
cluster production. 

This conclusion is naive for at least two reasons:\\
(i) 
Although it is in agreement with a general trend of
instanton suppression in high field, high temperature and high density
\cite{Shu_78_conf}, one has to look at each issue individually.
It was shown by \cite{KKL_inst} that dense saturated CGC of a $single$
nuclei does not actually suppress instantons\footnote{They did it by
an explicit
calculation: going into the rest frame of the nuclei can
convince the reader  much easier.}: this can only happen
in space-time region after the collision. 
On the  other hand, ``killed vacuum instantons'' will always
contribute,
but their number is limited and does not grow with energy.
\\
(ii) The saturation scale is not the scale of transverse
momenta/virtualities
which all the partons have. By its definition, it is only a place where 
the distribution over them changes, from a dilute to a saturated regime.
Thus a convolution over all momentum scales with realistic
instanton size distribution has to be evaluated, to establish
accurate energy dependence of the instanton-induced processes.
Such a calculation is not done yet. 
\subsection{Heavy Quarkonia Suppression} 
\label{sec_psi}
  In this subsection we consider one more possible signature of
cluster explosion:
 strong $J/\psi$
 ``ionization'' by the dipole excitation similar to those
responsible for photo-effect of ordinary atoms. The effect in QGP
context has been noticed
already in the very first paper discussing possible QGP signals \cite{Shu_80}.
Later Matsui and Satz  \cite{MS}  noticed that due to a Debye
 screening
in QGP
$J/\psi$ and other excited states of $\bar c c$ 
cannot  even exist inside QGP as a bound states. Since that time,
multiple literature on the subject discusses theory of $J/\psi$
suppression and its relation with experimental data from CERN NA38/50
 experiment. RHIC data are expected soon, and prediction range from
 near total suppression to even $J/\psi$ enhancement, due to
possible recombination of two charm quarks at the hadronization stage.

Without going into detailed discussion of current situation, we just
evaluate the probability to dissociate different quarkonia states
by the gluo-electric field of expanding shells resulting from
the cluster explosion. 

The cross section for a gluon (with momentum k)
to dissociate a quarkonium state, denoted as $\Phi$ is \cite{g_psi}
\be \label{eqn_diss_cross}
\sigma_{g\Phi}={2\pi \over 3} ({32 \over 3 })^2({M\over \epsilon})^{1/2}
{(k/\epsilon -1)^{3/2} \over M^2 (k/\epsilon)^5}
\ee
where $\Phi$ is assumed to be a Coulomb system made of quarks with mass
M and binding energy $\epsilon$.

Since the wall of the exploding cluster is a coherent
field, in principle one should go back to the original calculation
and calculate matrix element of excitation using time-dependent
perturbation
theory and appropriate time shape of the pulse. However,
since the maxima of the gluon spectra from cluster decay
 and the cross section  for the $J/\psi$ dissociation given above happen
to be nearly coincident, one can simply estimate it using the incoherent
cross section $enhanced$ by a coherence factor\footnote{The reader
  should not confuse this factor with a number of outgoing partons at
  the end. One may say that at the beginning  the cluster is made of
  $N_g\sim 6$ gluons, and at the end of 3 gluons plus 0-3 $\bar q q$.
} $N_g$=4-5.
(The fields of all gluons are the same and  add in a field strength,
not in intensity.)

If the quarkonium is at distance R from the original position
of the cluster, the probability to be dissociated
is 
\be P=N_g\sigma_{g\Phi}{N_g \over 4\pi R^2}\ee
where the unusual extra factor $N_g$ is the coherent enhancement.
Averaging it over the volume, assuming the space is divided into
spherical cells of radius $R_c$ around each cluster,
we get
\be 
P= N_g^2 n_c \sigma_{g\Phi}  R_c 
\ee
where $n_c=1/(4\pi R_c^3/3)$ is the cluster density.

One can see that for expected  $n_{clusters}\sim 1 fm^{-3}$ one gets
P$_{J/\psi}>1$,
which means that nearly all $J/\psi$ are expected to be dissociated.

The situation is different for $\Upsilon$ and its relatives.
First of all, the maximum of the cross section is down by about an
order
of magnitude. Second, 
the overlap between the gluon spectrum and the cross section 
(\ref{eqn_diss_cross}) is small: exact numbers also
depend  on (relatively uncertain) high energy tail $k> 1 \, GeV$ of
this
spectrum. As a result, an
 effective cross section is only $\sim 0.1 mb$ and the
probability of Upsilon excitation is just few percents.

Presumably remnants of ``classical glue'',
which according to calculations in \cite{KV} have higher  effective
temperature
$T_{eff}\sim 1 \, GeV$ are more effective in Upsilon dissociation.

\section{ Jet Quenching and the explosive edge}

\label{sec_intro_quench}
Jet quenching is  a kind of
  ``tomography''
of the excited system, created in high energy heavy ion collisions.
 It is based on the fact that even large-$p_t$ jets are 
partially absorbed\footnote{Or deformed: see discussion below on
  experimental strategies.}
during their passage through the system. Jet quenching is thus
a tools allowing us to get information
  about
very early stages of such conditions.
The so called {\em quenching factor} $Q(p_t)$ is defined as
 the observed number of high-$p_t$ hadrons
in AA collisions divided by the $expected$ number calculated
in the parton model, scaled from pp by a number of collisions.
For pions this factor is found to be at RHIC $Q(p_t)\sim .2, p_t>4\,
  GeV$,
smaller than expected from pQCD estimate.
What role in in is played bu the
   $initial$ state interaction -- the so called saturation or nuclear
shadowing of structure functions, and/or Cronin effects -- should be
clarified in the ongoing deuteron-Au run.

  Let us now turn to the theory of jet quenching.
In early papers on the subject \cite{early} only parton scattering   
in Quark-Gluon Plasma (QGP) were discussed.
 Account for radiation
losses \cite{Gyulassy_losses} and
 Landau-Pomeranchuck-Migdal (LPM) effect 
 \cite{Dok_etal} have finalized expectations based on the picture
of independent scattering centers. 

In recent work \cite{SZ_quench} a new picture has been proposed,
that instead of incoherent scattering points the matter is
actually made of coherent classical field, CGC or topological clusters.
The first $generic$ enhancement of the radiation
is due to coherence effect of classical glue: the field strength
are added, not intensities, of all gluons making the expanding wall.
(This would also work for classical glue created by a materialized
CGC.)

The second $specific$ enhancement is due to the fact that expanding
thin wall geometry maximizes both the field strength (compare to CGC
gluons which occupy all the space equally) and the probability
to collide with it. The last point may be especially important:
 it is hard for a jet to avoid 
the  foam-like structure of expanding shells.

We will not of course go into a rather involved calculations 
 of synchrotron-like radiation in a constant
gluo-magnetic  field performed in \cite{SZ_quench},
and just mention its main result for the quark energy loss

\be 
{\Delta E \over E}\sim 0.21\, 
\left({H  \over  0.2\, {\rm GeV}^2}\right)^{2/3}
\left({ 1 \, {\rm GeV} \over E}\right)^{1/3}\,\,.
\ee
where $H$ is the field in the shell.
The gluon loss scales with the pertinent color Casimir and is about 
twice larger. Convoluting such losses with the spectrum observed one finds
quenching factors of about one order of magnitude, which is in the
ballpark of observations.

  Let us further note that
two mechanisms of radiative energy losses discussed above, namely  (i) due to
multiple  uncorrelated scattering with small momentum transfer, and (ii)
one (or few) stronger scatterings can in principle be separated
experimentally. The difference between final states become apparent
when one asks a question: where the radiated energy go?
 Theoretically, when the emitting parton and its radiation
are separated (become incoherent), the radiation process is considered
completed. 
Experimentally however we cannot know anything about coherence and
just look at secondary hadrons, either individually or calorimetrically.
In the former case (i) most of the ``radiated'' energy still is in
forward
narrow cone. One may even  view this mechanism
of jet quenching as  a ``matter
modification of a fragmentation function'', with a conserved
total energy and
re-distributed hadron momenta. 
In the latter case (ii) (which 
we propose) the situation is different.
Radiated gluons  are scattered by a
strong 
field to large angels. Therefore, their energy 
is $not$ in the  narrow cone along the jet and
it {\em cannot be recovered}. This observation may give us a chance to
understand  which mechanism is responsible for jet quenching.

   Summarizing this section: the magnitude of the jet quenching
may be explained, by either CGC or cluster scenario. However,  
as we have already indicated in the list of ``RHIC puzzles'', the
most surprising observations are  rather
two other features: large azimuthal asymmetry and unusual
baryon/meson ratio at $p_t> 2 \, GeV$. 

Let me add a bit to discussion of $v_2$ at large $p_t$. It was found in my work
  \cite{Sh_v2_largept} that very strong jet quenching
  leads to some well-defined ``geometric limit'' of $v_2$ corresponding
to the surface emission,   but even this value falls
 below the experimental data. This  original conclusion,
  reflected
even in the title of this paper. However by the time the paper's
proofs were made,
STAR have performed additional studies of the 4-body cumulants \cite{STAR_v2} and
concluded that about 15 percent of the original $v_2$ was due to
non-flow
2-body correlations. That brought the data into agreement with the
geometric
limit for the sharp-edge nuclei. Further account for a ``nuclear
skin''
\cite{Casa_Shu} have however lowered the geometric limit as well, and
so
the contradiction between data and the geometrical limit is still there.

\begin{figure}[ht!]
\begin{center}
      \includegraphics[width=2.6in]{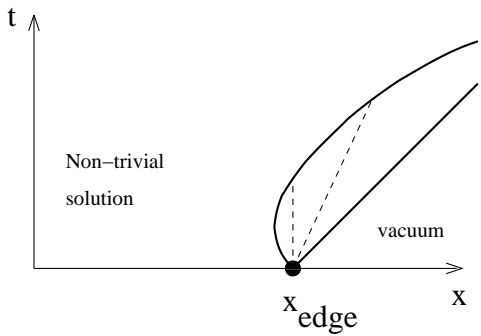}
\end{center}
  \caption{
   \label{fig_Riemann}
Schematic location of the Riemann
wave solution, on the plane the time t - the space x, between
the nontrivial solution of the hydro equations and empty space.
 Two solid line separating it are the light cone at the right and
 deflagration
front on the left. Rapidity and thermodynamical quantities are constant
on straight lines originating from the edge, shown by the dashed lines.
 }
\end{figure}


\subsection{The Explosive Edge: hydro treatment}

  Unusual phenomena at large $p_t$ observed at RHIC are clearly
  related with the surface of the excited system produced in the
  collision. This is clear already from
 the significant quenching of spectra, effectively excluding the
  central region,
but also from
large azimuthal asymmetry,
and absence of the backward jets in events with high-$p_t$ trigger.
Furthermore, the data on $v_2(p_t)$ show that hydro-generated
behavior at $p_t<2\, GeV$ joins very smoothly to a new (so far
unexplained) regime at $p_t >2\, GeV$. Also, it was observed that $v_2$ for
  baryons is somewhat larger than for mesons: natural for hydro regime
  it
seem to be true at higher $p_t=(2-6) \, GeV$ as well. All this hints to a
possibility  that the basic physics in
the hydro domain and that at higher $p_t$ are somewhat related.  

There is however a generic problem, preventing us from combining hydro and
jet physics into any combined model: 
their quite different {\em time scales}. Indeed, 
jets move with a speed of light, and cannot be emitted too far from
the surface: so
 escaping jet fragments  can only interact with
matter
for a short time, $t_{quenching}\sim 1 \, fm/c$.
Hydro flow, on the other hand, need longer time to be developed,
about  $t_{hydro}\sim 10\, fm/c$ for radial and about half of that for elliptic flow.

The way out of this dilemma proposed in this section
is to focus on  the specific collective phenomena
which may develop during the short time   $t_{quenching}$ at the edge
of the fireball. We will do so first in the hydrodynamical context,
where the analytic answers can be obtained. This however can only be
used as qualitative indications since the
size of the system (or the particle
number) is not large enough to justify
this approach quantitatively,  In the second
subsection
we suggest that multiple explosion of clusters may also
``jump-start'' collective phenomena at the surface.
 

Let us start with the generic problem, of an explosion of a system
which has a sharp edge. The main point to make here on the onset is
that phenomena near the edge  basically drive explosions in 1d, which
are much more robust that those in 2d and 3d.

Do we have an edge in heavy ion collisions? 
Indeed, if one ignores ``skin'' of nuclei
and treat them as drops of nuclear matter with a sharp spherical
surface, the almond-shaped overlap region of two nuclei colliding with
the impact parameter $b$ will have a sharp edge at which the energy 
(also particle and entropy) density vanishes as
$\epsilon(x) \sim \sqrt{x-x_{edge}}$. From hydrodynamical
equation it then follows that the acceleration at this point is  formally
infinite\footnote{If the edge is regulated by the usual skin with a
  Fermi-type
distribution, the acceleration is finite and constant at large x,
but inversely  proportional to small skin width. }

\be {\partial v \over \partial t}\approx {1\over \epsilon+p}{\partial
  p\over \partial x}\sim {1 \over x-x_{edge}}
   \ee
 
  It is well known how the problem is resolved, at least
  mathematically,
in the framework of hydrodynamical equations. 
The
singularities in the {\em initial conditions} are extensively
discussed in textbooks such as
\cite{LL_hydro}: let me remind the main idea only.
 Although hydro admits discontinuities (such as shocks),
they always come with certain conditions on them (such as Gugonio/Taub adiabats etc)
which the initial conditions in general do not obey. Therefore,
a generic situation is that a singular point in the initial conditions
opens up into a whole region of special solution, 
separated by 2 
discontinuities from other regions.

A singularity of the type we discuss opens into a region filled with
the so called Riemann wave in which rapidity and thermodynamical
quantities can be viewed as functions of each other.
The region of this solution is schematically shown in
 Fig.\ref{fig_Riemann}, it is separated from vacuum by a light cone
$t=x-x_{edge}$ at which matter content vanishes, and from 
the non-trivial hydro solution on the left
by a  deflagration front.
 (The latter curves at later time because the developed hydro rapidity
reaches values larger than that of the deflagration.)
  
The Riemann wave solution itself in relativistic flow is discussed 
in \cite{LL_hydro}, problem 1 to section 134. It can be summarized by
two
equations
\be y=\int {c_s d\epsilon \over p+\epsilon}  \ee
\be  x-x_{edge}=t* tanh(y-y_s) \ee
where $y,y_s$ is matter and sound rapidity, $v=tanh(y),c_s=tanh(y_s)$. 
$\epsilon$ is energy density, $p$ is pressure and $c_s^2=dp/d\epsilon$
is squared sound velocity. If $p,\epsilon$ is a function of one
variable (e.g. temperature) the meaning of the integral in the former
equation is clear. If there are more conserved quantities and chemical
potentials, it should be taken along the adiabatic 
path in the phase diagram the systems takes while cooling down.
In particular, for QGP or resonance gas with a simple 
EoS $c_s^2=const$ this first equation establishes
a very simple
 direct relation between particle/entropy density $\sim T^{1/c_s^2}$
 and rapidity, namely 
\be {n\over n_0}=exp(-y/c_s) \ee
each of which holds at the whole line originating from the edge.
The light cone corresponds to $y\rightarrow\infty,n\rightarrow 0$,
$n_0$ is the density at rapidity zero, etc. Of course, exponent
in rapidity is a 
power law in momenta/energies, and its index is rather unusual. For
QGP Eos it gives
$1/p^{\sqrt{3}}$, with a somewhat larger power 2-3
if one uses more appropriate EoS of the resonance gas.
 However in all cases the spectrum generated by the Riemann wave is much
harder than the observed particle spectrum, which has larger effective
power
of about 10-12.

Let us now discuss limitations on hydrodynamics stemming from the fact
that the particle multiplicity in our problem is not  like
 $10^{23}$ as in air/water but rather limited. At central AuAu
collisions at RHIC at rapidity $y\sim 0$ one actually finds\footnote{We mean
 all kinds of hadrons together, charged and neutral. The partons
are not directly observed, of course, but believed to be about the same
to justify the approximately adiabatic expansion 
which works well in hydro applications and confirmed by cascades.}
\be {dN_{hadrons} \over d\eta d\phi} \sim  {dN_{partons} \over d\eta d\phi} 
\sim 150 \ee
where $\eta,\phi$ are pseudo-rapidity and azimuthal angle.
This means that when we go about 2 orders of magnitude down the
spectra
we reach of only about 1 particle per unit solid angle, and should stop using
hydrodynamics. More elaborate estimates using viscosity corrections
\cite{Teaney_visc} lead to the same conclusion.

This simple idea put a cutoff of hydro
to $p_t\sim 1.7-2\, GeV$, which is indeed where deviations from hydro
predictions are observed. 
It also indicate to us that the above discussion of the Riemann wave
can only be used as a qualitative indication, and its comparison
to data is not really justified.

\begin{figure}[t]
\centering
\includegraphics[width=7cm]{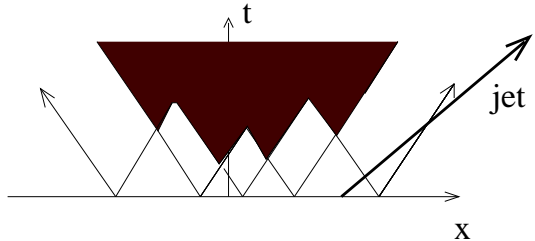}
\caption{\label{fig_edge} 
 Schematic picture of initial stage of the heavy ion collision,
on the plane transverse coordinate x - time t. At t=0 several clusters
 are produced which promptly 
 explode into  empty spherical shells,
shown by two divergent lines. The 
 decoherent quarks and gluons -- QGP shown as the shaded region--
 is produced in secondary collisions
of these shells, but partons at the edge fly away.
}
\end{figure}

\subsection{Exploding Clusters at the edge}
 We have concluded at the end of the previous section, that hydro
can provide only qualitative ideas about matter evolution at the edge
of the system, and thus some microscopic approach is needed.

This is of course very much limited by our poor knowledge of the
matter conditions at the early time. However the generic idea -- that
particles at the edge of the system experience collisions with
a stream of particles from one side only and are therefore
accelerated more rapidly than any others -- holds in one form or
another for any models under
consideration.

In a perturbative mini-jet scenario with a modest re-scattering one
simply
would find a fraction of outward moving mini-jets which never
re-scattered,
but fragment into secondary hadrons independently. Naturally, those
would spatially separate from those which did re-scatter, and create
thin collisionless ``atmosphere'' receding from the fireball with
(transverse) rapidity $y_t\sim 1$.

In a CGC picture one should study the behavior of the 
classical Yang-Mills equations which starts with the initial conditions
having a sharp edge. Although it is not yet done, by analogy to discussion
of hydro equation in the preceding section, one may expect the edge
also to act as a focal plane of an explosive expansion into vacuum.
Moreover, in terms of effective pressure the non-linear term in YM
equation is even stronger than in the EoS of QGP, so more dramatic
phenomena can be expected.

Finally, in the scenario advocated in this paper, namely
on multiple production of the topological clusters, one get a picture
shown schematically in  Figure
\ref{fig_edge}.
The message displayed there is very simple:
although most expanding spherical
 shells from the cluster explosion 
collide and produce QGP, as discussed above, the part of those shells
going from the edge toward the empty space cannot be not stopped in
this way. 
 
So, if in all pictures of the initial stage one finds partons
receding from the system edge outward, it is obviously interesting to know
what happens with them?  The answer to this question
depends on their density. In pp collisions, with a couple of mini-jets
or a
single cluster produced, confining flux tubes 
 would slow down their propagation and drain their energy into
flux tubes fragmentation (hadronization a la Lund model). 

We would argue that the situation should be different if the outward
moving partons are sufficiently numerous to neutralize each other
in color and thus fly away {\em without strings attached}. In other
words, we suggest that parton fragmentation in AA collisions is
different from independent one observed in $e+e-$ and $pp$ collisions.

We will further suggest that even fragmentation of high-$p_t$ jets
should be affected, as they fly through the comoving matter originated
from
the explosive edge, see  Figure
\ref{fig_edge}. A number of detailed models 
of such kind are under numerical investigation now, with
results to be reported elsewhere \cite{Casa_Shu}. Hopefully those
provide
some insight into the puzzles of large $v_2$ and high baryon content.

\section{Event-by-event Fluctuations}
\label{sec_fluct}
First, a brief history.
This approach was first proposed \cite{Stodolsky_ES} as a tool to test
statistical methods and extract more accurate data about freeze-out.
More radical ideas of its applications included a search for the QCD
tri-critical point \cite{SRS} and even for quark-gluon plasma
\cite{AHM_JK}.

Experimentally, it has been pioneered by NA49 experiment at CERN \cite{NA49}
which found that non-statistical event-by-event fluctuations in mean $p_t$
are basically absent,
if measurements are made for rather forward rapidities. Late data by
this group, as well as by CERES experiment, have found dynamical
effect at mid-rapidity roughly consistent with standard resonance
production: no trace so far neither of the tri-critical point nor of
survived
 QGP fluctuations.

The first STAR preliminary data on mean $p_t$ fluctuations
have been discussed in a number of meetings by  
T.Trainor  and G.Roland, but unfortunately 
no publications are so far available.

Cluster production  induce 
 {\em event-by-event
fluctuations} because their 
 number
 is significantly smaller than the number of particles. The
estimates thus follows the ordinary statistical arguments, assuming
separate cluster production is independent. The
 original presentation of this idea
 has been made at the 2001 CERN workshop \cite{fluct_hif}.
Discussion of the elliptic flow or $v_2$ fluctuations 
of the same origin is discussed in a separate paper 
\cite{Mrow_Shu}.

The relative fluctuation in number of clusters
is 
\be {\delta dN_{clust}/dy \over dN_{clust}/dy} \sim ({1 \over
dN_{clust}/dy})^{1/2}
\ee 
For central AuAu collisions at RHIC this fluctuation is about .1.
Assuming the fraction of in total parton multiplicity  coming from clusters, $f_{clust}\approx
  1/2$, and
that the other component is fluctuating much less, we
then
expect the observed particle density per unit rapidity $dN/dy$ 
at mid-rapidity of RHIC AuAu central collisions to fluctuate by 
about by 0.05. (This of course can be directly tested experimentally).

The next step is  to estimate the expected fluctuations in hydro
expansion velocity:
\be 
{\delta v_t \over v_t}={\delta dN/dy \over dN/dy} 
 {\partial log (v_t) \over \partial log (dN/dy)}
\ee
The log-log 
derivative,
also called index, depends on the EOS and can be calculated from hydrodynamics. 
Simple estimate follows from a comparison of mean flow
at SPS and RHIC (or at two RHIC
energies). At $\sqrt{s}$ changing from 17 to 130 GeV the charge in hadron
multiplicity $dN/dy$ is about 1.5 while the flow velocity
changed only from .55 to about .65:
thus the log-log derivative
is about .4, so relative fluctuation
in the velocity is estimated to be about 0.02. 

The so called $m_t$ slopes, or effective
temperatures $T_{slope}$,
 contains $v_T$ via the so called {\em blue
shift factor}
\be T_{slope}=T_{true}\sqrt{ \frac{1+v_{t}}{1-v_{t}} }  \ee
Differentiating this over $v_t$
we finally get the estimate for the  ``temperature fluctuation''  at RHIC 
relative to SPS (where cluster production is assumed to be negligible)
\be {\delta T_{slope} \over T_{slope}}={\delta v_t \over v_t}{v_t \over 1-v^2_t}  \ee
or about 0.01 in central AuAu collisions.
A relative increase in fluctuation in STAR relative to NA49 is
indeed of the magnitude corresponding to about 1 per cent
``slope fluctuations''
for the central heavy ion collisions. 

Another way to estimate the log-log derivative or index is directly from the
hydrodynamical calculations.. Fig.14 of the second 
ref. of \cite{hydro} show dependence of $<m_t>-m$ for pions and nucleons 
on the input multiplicity $dN/dy$. Using these results for two highest values,
$dN/dy=466, 669$, we get that the corresponding log-log derivative 
\be P_h={\partial log T_{slope} \over\partial log dN/dy} \ee
are  $P_\pi= .37$ for pions and $P_N=.63$ for nucleons. Together with
\be {\delta T_{slope} \over T_{slope}}=P_h {\delta dN/dy \over dN/dy} 
\ee
The resulting prediction is again about 1 percent fluctuation of slope for pions but about 2 
percent for nucleons. (This difference is natural since nucleons are more sensitive
to flow in general.) 

 More direct tests of whether event-by-event fluctuations of mean
$p_t$ are indeed due to the $p_t$-slope fluctuations,
or something else, can be
done experimentally using complete {\em two-body correlation
function}.
Simple summary of the idea can be expressed as follows:
Slope fluctuation produce $positive correlation$ when two transverse
 momenta are either both
small or both large, but a $negative$ one
 if one momentum is large and one small.
Preliminary indications are that it is indeed the case: we are waiting
for data to get finalized and public, to test  it further.

Finally, a comment on other possible kind of event-by-event
fluctuations, those related with the {\em non-zero topology} of the
clusters,
which (again with relative probability $\sim1/\sqrt{N_{clusters}}$
which is large compared to statistical fluctuations
$\sim1/\sqrt{N_{particles}}$  )
would produce an observable
 dis-balance of {\em left-versus-right quarks}. Like in
electroweak theory in Big Bang before electroweak transition, 
we think that thermal rate in QGP should be
 big enough 
to erase this effect.

\section{Summary and Discussion}
\label{sec_conclusions}
We have discussed in this work multiple consequences of the idea,
that quantum mechanical excitation from $under$ the
QCD topological barrier can be an important component
of high energy heavy ion collisions, and at RHIC energies
($\sqrt{s}\sim$ hundreds of GeV) in particular.
We thus proposed that more traditional views of the initial stage
of such collisions be supplemented by
 a ``firework'' of mini-explosions of 
topological clusters, the remnants
of vacuum instantons.  Both their shape and further explosive behavior
are determined from classical YM and is under control, 
numerically and analytically.
We pointed out that the low scale of the thermal masses
can be used to justify application of classical Yang-Mills equations,
 as clusters
 explodes into QGP. Each cluster
 leads to  production of about 3 gluons+(0-3)$\bar q q$. 
 Prompt production of quarks by this mechanism
is especially unique, and may leads toward QGP, although
not quite in the equilibrium g/q ratios.

 We made rough estimates of  the number of clusters
produced
based on parton-parton cross section obtained phenomenologically
(more work on its direct calculation is still needed) and found
its upper limit to be comparable to amount of entropy actually produced. 
 Other mechanism (such as Color Glass Condensate advocated by
McLerran-Venugopalan
model) also may produce a similar entropy but with much higher
temperature. A combined entropy and energy of both components
roughly correspond to initial properties of equilibrated QGP, known
from hydro applications.
 
  The signature of the topological clusters is thus
prompt production of quark-anti-quark
pairs. 
 With lepton and/or photons from early stages
to be eventually observed, 
one would be eventually be able to probe the quark/gluon
ratio at early stages
experimentally. Another place where prompt quark production
is needed is to explain large baryonic component at high $p_t$.
Although we have not proposed a specific model for it, we argued that
the existence of the ``explosive edge'' containing quark pairs is
likely
to be related with their origin. 

We further argued that cluster production is expected
to become sub-leading at energies much higher than RHIC, at LHC,
$provided$  the saturation scale there is indeed  increased well above
the instanton scale. 

 We further speculated
that cluster production can help us with some other RHIC puzzles,
such as enhanced event-by-event fluctuations and jet quenching. 
These estimates are  to be compared with RHIC data, as they
will become more detailed and accurate.

{\bf Acknowledgments}
Let me thank my collaborators G.Carter,J.Casalderry, R.Janik,
D.Ostrovsky and  I. Zahed: this paper is in many ways a summary of
ideas
which have been developed in the course of joint work  with them.
I also would like to thank D.Kharzeev,L.McLerran and R.Venugopalan
 for useful discussion of their papers.
This work is partially supported by the US-DOE grant
No. DE-FG02-88ER40388. The main part of it has been written in ITP, Santa
Barbara provided support under NSF Grant PHY99-07949. 
Kind hospitality and very interesting discussions with people in ITP  and 
especially with David Gross is acknowledged.

\end{document}